\documentclass[aps,showpacs,onecolumn,superscriptaddress]{revtex4}
\usepackage{epsfig,verbatim}
\begin{document}

\font\Bbb =msbm10  scaled \magstephalf \def\id{{\hbox{\Bbb I}}}
\newcommand{\ket}[1]{\big| \, #1 \big\rangle}
\newcommand{\bra}[1]{\left \langle #1 \, \right |}
\newcommand{\proj}[1]{\ket{#1}\bra{#1}}
\newcommand{\braket}[2]{\langle\, #1\,|\,#2\,\rangle}
\newcommand{\half}{\mbox{$\textstyle \frac{1}{2}$}}
\def\opone{\leavevmode\hbox{\small1\kern-3.8pt\normalsize1}}
\newcommand{\tr}[1]{\mbox{Tr} \, #1 }
\def\emph#1{{\it #1}}
\def\textbf#1{{\bf #1}}
\def\textrm#1{{\rm #1}}
\newcommand{\bea}{\begin{eqnarray}}
\newcommand{\eea}{\end{eqnarray}}
\newcommand{\beq}{\begin{equation}}
\newcommand{\eeq}{\end{equation}}

\baselineskip=21pt

\title{Exploration of CPT violation via time-dependent geometric quantities embedded in neutrino oscillation through fluctuating
matter}

\author{Zisheng Wang}
\thanks{zishengwang@yahoo.com}
\affiliation{College of Physics and Communication Electronics,
Jiangxi Normal university, Nanchang 330022, P. R. China}
\affiliation{Institute of Applied Physics and Materials Engineering,
Faculty of Science and Technology, University of Macau, Macao SAR,
China}
\author{Hui Pan}
\thanks{huipan@umac.mo}
\affiliation{Institute of Applied Physics and Materials Engineering,
Faculty of Science and Technology, University of Macau, Macao SAR,
China}

\begin{abstract}

We propose a new approach to explore CPT violation of neutrino oscillations through a fluctuating matter based on
time-dependent geometric quantities. By mapping the neutrino oscillations onto a Poincar\'e sphere structure, we obtain an analytic solution
of master equation and further define the geometric quantities, i.e., radius of Poincar\'e sphere and geometric phase. We find that the mixing process between electron and muon neutrinos can be described by the radius of Poincar\'e sphere that depends on the intrinsic CP-violating angle. Such a radius reveals a dynamic mechanism of CPT-violation, i.e., both spontaneous symmetry breaking and Majorana-Dirac neutrino confusion. We show that the time-dependent geometric phase can be used to find the neutrino nature and observe the CPT-violation because it is strongly enhanced under the neutrino propagation. We further show that the
time-dependent geometric phase can be easily detected by simulating the neutrino oscillation based on
fluctuating magnetic fields in nuclear magnetic resonance, which makes the experimental observation of CPT-violation possible in the neutrino mixing and oscillations.
\end{abstract}

\maketitle

\section{Introduce}

Neutrino mixing and oscillations are important to investigate
new physics beyond the Standard Model of elementary particle
physics, and also involve in hot issues on both astro-particle physics and
cosmology \cite{{1},{2},{3}}. Experimentally and
theoretically, a set of important and fundamental problems, such as
the neutrino mass, the nature of the Dirac vs Majorana neutrino or Majorana-Dirac confusion theorem \cite{4},
and the validity of CPT symmetry, has been under the debate \cite{{5},{6},{7},{8},{9}}. The interference of
the neutrino oscillations \cite{{10},{11}} can be used to test the CPT symmetry and the neutrino nature based effectively on the geometric phase, which provides an unconventional approach to probe CPT-violation beside the neutron electric dipole moment \cite{12}.

The evolution of neutrinos involved the weak interactions leads to the CP-violation in a given flavor space,
where the extrinsic CP-violating phases can mimic the
characteristics of intrinsic CP-violating phases in the leptonic mixing
matrix \cite{{13},{13},{14},{15},{16}}.

The neutrino oscillations observed
so far can be explained in terms of three flavor space, i.e., active electron neutrino
 $\nu_e$, muon neutrino $\nu_\mu$ and $\tau$ neutrino $\nu_\tau$, where the extrinsic CP violation is disentangled from the intrinsic one by the CP-violating observable \cite{17}. An alternative method to treat neutrino oscillations has attracted extensive interests in terms of analogizing with simple and intuitive
Rabi-oscillations in a two-flavor space \cite{{1},{18}}, where the extrinsic CP-violating phases can
exactly appear in the two-level Hamiltonian.

In principle, such a CP-violating phase can be observed in neutrinoless double beta decay (NDBD) in terms of the transition probability, but it is an extremely low efficiency of direct neutrino measurements so as to have not fully solve this problem in the experiments up to now.

The characteristics of mixed neutrino evolution can be
recognized by the geometric phase \cite{{19},{20}}. Especially, such a geometric phase can be observed in the experiments. In the neutrino mixing and oscillations, the neutrino system interacts irreversibly with its
surrounding environment \cite{{21},{22}}, resulting in statistical mixtures
of quantum superpositions. To date, most studies of geometric phase have been focused on the pure state by using quantum mechanics in terms of a closed system \cite{11} or have not included the two-state mixing effects in the open system \cite{{20},{23},{24}}, which are unsuitable for neutrino mixing and oscillations. It is, therefore, necessary to find out the mechanism of flavor neutrino mixing and oscillations in the geometric phase just like the mixed state of open system.

In this work, we map the two-flavor neutrino oscillations onto a three-dimensional Poincar\'e sphere structure. We find that the time-dependent radius of Poincar\'e sphere can describe neutrino mixture degree of freedom and reveal a dynamic mechanism of CPT violation. We further investigate the relations between the time-dependent geometric quantities and CPT violation. Finally, we propose a new approach to detect the geometric phase with the CPT-violating effects by quantum simulation of physical fluctuating fields.

\section{Hamiltonian with a CP-violating phase}

Let us consider solar neutrinos with small square-mass
difference $\Delta m^2_{21}=m^2_2 - m^2_1$ between the neutrinos $\nu_\mu$ and
 $\nu_e$, where the large
squared-mass ones $\Delta m^2_{31}$ and $\Delta m^2_{32}$ between $\nu_\tau$ and
$\nu_e$ and between $\nu_\tau$ and
$\nu_\mu$ are averaged out \cite{{1},{2}}. Under the mass scale dominant
approximation and in the ultra-relativistic limit $p \approx E$, the Hamiltonian ${\cal H}$ with an intrinsic CP-violating phase $\phi$ for two flavor neutrino oscillations in medium can be
expressed as \cite{{1},{5}}

\begin{eqnarray}
{\cal H} &=& \left( E + \frac{m^2_1 + m^2_2}{4E} +
\frac{V_0}{2}\right){\cal I}_{2\times 2} +
\frac{1}{2}\left(\begin{array}{cc} V_0-\frac{\Delta m^2_{21}}{2E}\cos2\theta & \frac{\Delta m^2_{21}}{2E}
e^{-i\phi}\sin2\theta \\ \frac{\Delta m^2_{21}}{2E} e^{i\phi}\sin2\theta  &
- V_0 + \frac{\Delta m^2_{21}}{2E}\cos2\theta
\end{array} \right),
\end{eqnarray}
where $\theta$ is a neutrino mixing angle in vacuum and
$V_0=\sqrt{2}G_Fn_e\cos^2\theta_{13}$ is a matter potential with the Fermi weak coupling constant $G_F$, the
electron density $n_e$ in the medium, and the oscillation parameter
$0.953<\cos^2\theta_{13}\le 1$ under $3\sigma$ bound. In terms of Mikheyev-Smirnov-Wolfenstein effect, the CP asymmetric matter potential $V_0$ can
enhance and suppress the oscillations in the neutrino and antineutrino channels, respectively.

Since ${\cal I}_{2\times 2}$ is a $2\times 2$ identity matrix,
the first term with non-zero trace on the right of Eq. (1) adds only an unimportant overall
phase factor to the time-evolving state in the neutrino mixing and oscillations. The second term can be divided
into diagonal and nondiagonal parts, which can be expanded in terms
of Pauli matrices (i.e., $\sigma_z, \sigma_x$ and $\sigma_y$). Since
only $\sigma_y$ changes sign under the charge conjugation and parity
(CP) transformation, $\phi$ is an intrinsic CP-violating (or
Majorana) phase. For the Dirac neutrino, $\phi$ can be eliminated by
a $U(1)$ gauge transformation. In
contrast, the rephasing of the left-chiral massive neutrino field is
not possible for the Majorana neutrino because the mass term of the Lagrangian is not invariant
under the gauge transformation.

In the Schr\"odinger picture, the amplitude $\Psi_{\alpha\beta}(t)=\langle\nu_\beta (0)\mid\nu_\alpha (t)\rangle$ of neutrino transition $\nu_\alpha\rightarrow\nu_\beta (\alpha=e, \nu=e,\mu)$ is evolved by \cite{1}

\begin{eqnarray}
i\frac{d}{dt}\Psi_{\alpha\beta}(t) = {\cal H}\Psi_{\alpha\beta}(t),
\end{eqnarray}
in the units $\hbar=1$, where the overall phase factor from the first term of Hamiltonian ${\cal H}$ (See Eq. (1)) can be eliminated by the phase shift,
\begin{eqnarray}
\Psi_{\alpha\beta}(t) = \psi_{\alpha\beta}(t)\exp{\left(-i (E + \frac{m^2_1 + m^2_2}{4E} +
\frac{V_0}{2})t\right)}.
\end{eqnarray}
Thus the relevant evolution equation can be expressed in terms of the matrix form, i.e.,
\begin{eqnarray}
i\frac{d}{dt} \left(\begin{array}{c} \psi_{ee}(t) \\
 \psi_{e\mu}(t)
\end{array} \right) = \frac{1}{4E}\left(\begin{array}{cc} -\Delta m^2_{21}\cos2\theta + A_{cc} & \Delta m^2_{21}\sin2\theta e^{-i\phi}\\
\Delta m^2_{21}\sin2\theta e^{i\phi} & \Delta m^2_{21}\cos2\theta - A_{cc}
\end{array} \right)
\left(\begin{array}{c} \psi_{ee}(t) \\
 \psi_{e\mu}(t)
\end{array} \right)
,
\end{eqnarray}
where $A_{cc}=2EV_0$. In Eq. (4), $\psi_{ee}(t)$ and $\psi_{e\mu}(t)$ are the transitive amplitudes of $\nu_e\rightarrow\nu_e$  and $\nu_e\rightarrow\nu_\mu$, respectively. The evolution equation (4) is a Schr\"odinger-like equation with effective Hamiltonian matrix in the flavor basis, which can be simplified by a  unitary transformation,
\begin{eqnarray}
\left(\begin{array}{c} \psi_{ee}(t) \\
 \psi_{e\mu}(t)
\end{array} \right)= U_M \left(\begin{array}{c} \phi_{e1}(t) \\
 \phi_{e2}(t)
\end{array} \right),
\end{eqnarray}
and

\begin{eqnarray}
{\cal H}_M = U^\dag_M\frac{1}{4E}\left(\begin{array}{cc} -\Delta m^2_{21}\cos2\theta + A_{cc} & \Delta m^2_{21}\sin2\theta e^{-i\phi}\\
\Delta m^2_{21}\sin2\theta e^{i\phi} & \Delta m^2_{21}\cos2\theta - A_{cc}
\end{array} \right) U_M = \frac{1}{4E}\left(\begin{array}{cc} -\Delta m^2_{M} & 0\\
0 & \Delta m^2_{M}
\end{array} \right),
\end{eqnarray}
where ${\cal H}_M$ is called as an effective Hamiltonian matrix in matter in terms of the mass basis and the effective squared-mass difference in matter is given by
\begin{eqnarray}
\Delta m^2_{M} = \sqrt{(\Delta m^2_{21}\cos2\theta - A_{cc})^2 + (\Delta m^2_{21}\sin2\theta)^2}
.
\end{eqnarray}

The unitary matrix in Eqs. (5) and (6) can be expressed as
\begin{eqnarray}
U_M = \left(\begin{array}{cc} \cos\vartheta_M & \sin\vartheta_M e^{-i\phi}\\
-\sin\vartheta_M e^{i\phi} & \cos\vartheta_M
\end{array} \right)
,
\end{eqnarray}
which is an effective mixing matrix with the effective mixing angle $\vartheta_M$ and the CP-violating phase in matter. The mixing angle $\vartheta_M$ in matter is written as
\begin{eqnarray}
\tan 2\vartheta_M = \frac{\tan2\theta}{1 - \frac{A_{cc}}{\Delta m^2_{21}\cos2\theta}},
\end{eqnarray}
where there is a resonance when $A_{cc}=\Delta m^2_{21}\cos2\theta$. At the resonance the effective mixing angle is $\vartheta=\pi/2$, i.e., the mixing is maximal, leading to the possibility of total transitions between the two flavors if the resonance region is wide enough.

Inserting Eqs. (5) and (6) into Eq. (4), one has
\begin{eqnarray}
i\frac{d}{dt} \left(\begin{array}{c} \phi_{e1}(t) \\
 \phi_{e2}(t)
\end{array} \right) = \frac{1}{4E}\left(\begin{array}{cc} -\Delta m^2_{M} & 0\\
0 & \Delta m^2_{M}
\end{array} \right)
\left(\begin{array}{c} \phi_{e1}(t) \\
 \phi_{e2}(t)
\end{array} \right)
,
\end{eqnarray}
with the initial conditions,
\begin{eqnarray}
\left(\begin{array}{c} \phi_{e1}(0) \\
 \phi_{e2}(0)
\end{array} \right) =
U_M^\dag\left(\begin{array}{c} \psi_{ee}(0) \\
 \psi_{e\mu}(0)\end{array} \right) = \left(\begin{array}{cc} \cos\vartheta_M & -\sin\vartheta_M e^{-i\phi}\\
\sin\vartheta_M e^{i\phi} & \cos\vartheta_M
\end{array} \right)
\left(\begin{array}{c} 1 \\
 0
\end{array} \right) = \left(\begin{array}{c} \cos\vartheta_M \\
 e^{i\phi}\sin\vartheta_M
\end{array} \right)
.
\end{eqnarray}

Eq. (10) is a decoupled evolution of the amplitudes of effective massive neutrinos in matter. Under the initial condition (11), its solution is direct. We find
\begin{eqnarray}
\left(\begin{array}{c} \phi_{e1}(t) \\
 \phi_{e2}(t)\end{array} \right)  = \left(\begin{array}{c} e^{i\frac{\Delta m^2_M}{4E}t}\cos\vartheta_M \\
 e^{i(\phi - \frac{\Delta m^2_M}{4E}t)}\sin\vartheta_M
\end{array} \right)
,
\end{eqnarray}
and
\begin{eqnarray}
\left(\begin{array}{c} \psi_{ee}(t) \\
 \psi_{e\mu}(t)\end{array} \right) = \left(\begin{array}{c} \cos^2\vartheta_M e^{i\frac{\Delta m^2_M}{4E}t} + \sin^2\vartheta_M e^{-i\frac{\Delta m^2_M}{4E}t} \\
 \frac{1}{2}e^{i\phi}\sin2\vartheta (e^{-i\frac{\Delta m^2_M}{4E}t} - e^{i\frac{\Delta m^2_M}{4E}t})
\end{array} \right)
,
\end{eqnarray}
which leads to the transition probabilities,
\begin{eqnarray}
P_{\nu_e\rightarrow \nu_\mu}(t) = |\Psi_{e\mu}(t)|^2 = |\psi_{e\mu}(t)|^2 = \sin^2 2\vartheta_M \sin^2 \left(\frac{\Delta m^2_M t}{4E}\right), P_{\nu_e\rightarrow \nu_e}(t) = 1 - P_{\nu_e\rightarrow \nu_\mu}(t)
,
\end{eqnarray}
which the CP-violating (or
Majorana) phase $\phi$ vanishes in the transition probability with the same structure as the two-neutrino transition probability in vacuum. As pointed by Giunti \cite{25}, such a Majorana phase of the neutrino mixing matrix cannot have any effect in the neutrino oscillations in the conventional
Hilbert space.

\section{Neutrino mixing and oscillations in dissipative matter}

It is known that the CP-violating effects can not be described in the two-flavor neutrino
mixed states in terms of the conventional
Hilbert space. In reality, no quantum system is completely isolated from its surroundings, so the neutrino system is open to some extent, causing dissipation in the quantum system.
When the neutrino propagates through a dissipative matter, the neutrino mixing and oscillations should be extended to  an open quantum system including the interactions between the neutrino and its external environment.

In this situation, a density matrix $\rho(t)$ needs
to be introduced in order to describe the two-flavor neutrino mixed
states with the synthesized properties, i.e.,  hermitian, positive
operators with non-negative eigenvalues, and unit
trace. The dynamic evolution of neutrino mixing and oscillations can be described by the Lindblad master equation \cite{26}, i.e.,

\begin{eqnarray}
\frac{d}{dt}\rho(t) = -i[{\cal H},\rho(t)] + {\cal L}\rho,
\end{eqnarray}
in the units $\hbar=1$, where the Lindblad superoperator is given by
\begin{eqnarray}
{\cal L}\rho = \frac{1}{2} \sum_{i,j=x,y,z}c_{ij}( [\sigma_i\rho,
\sigma_j] + [\sigma_i,\rho \sigma_j] ),
\end{eqnarray}
where the summation runs to $N^2-1 =2^2-1$ for the two-dimensional quantum system  and therefore includes all possible decay ways with the constant coefficients $c_{ij}\ge 0$ due to the interaction between the neutrinos and dissipative environment.

Differently from the Schr\"odinger picture in Eq. (2), the master equation (15) incorporates decoherent effects due to energy loss ($i\ne j$) or loss of pure phase information without energy ($i=j=z$). Thus Eq. (15) essentially goes beyond the standard Hamiltonian dynamics. The first term on the right of Eq. (15) is a usual Schr\"odinger term but includes the CP-violating effect as shown in Eq. (1). The second term leads to time-irreversibility. Therefore, we can explore the CPT violating effect in terms of the master equation (15).

At initial time ($t=0$), the two flavor states, the electron neutrino $\mid\nu_e\rangle$ and the muon
neutrino $\mid\nu_\mu\rangle$, are described by the pure states
and can be represented by
\begin{eqnarray}
\mid\nu_e(0)\rangle = \left(\begin{array}{c} \cos\theta \\
 e^{i\phi}\sin\theta
\end{array} \right), \mid\nu_\mu(0)\rangle = \left(\begin{array}{c} \sin\theta \\
- e^{i\phi}\cos\theta
\end{array} \right),
\end{eqnarray}
with the corresponding density matrices,
\begin{eqnarray}
\rho_e(0) = \left(\begin{array}{cc} \cos^2\theta & \frac{1}{2}e^{-i\phi}\sin2\theta\\
 \frac{1}{2}e^{i\phi}\sin2\theta & \sin^2\theta
\end{array} \right), \rho_\mu(0) = \left(\begin{array}{cc} \sin^2\theta & -\frac{1}{2}e^{-i\phi}\sin2\theta \\
-\frac{1}{2}e^{i\phi}\sin2\theta & \cos^2\theta
\end{array} \right),
\end{eqnarray}
which can be reexpressed by the Pauli matrices, i.e.,
$\rho_{e}(0) = \frac{1}{2}({\cal I}_{2\times 2} +
\sin(2\theta)\cos\phi\sigma_x + \sin(2\theta)\sin\phi\sigma_y +
\cos(2\theta)\sigma_z)$ and $\rho_{\mu}(0) = \frac{1}{2}({\cal
I}_{2\times 2} - \sin(2\theta)\cos\phi\sigma_x -
\sin(2\theta)\sin\phi\sigma_y - \cos(2\theta)\sigma_z)$. Since the $\sigma_y$'s coefficients are not equal to zero, the initial density matrices carry with the CP-violating messages.

\section{Poincar\'e sphere structure}

A geometric representation of neutrino mixing and oscillations is an
effective approach to understand and analyze the dynamic evolution
of neutrino system \cite{{27},{28}}. Therefore we map the neutrino oscillations in the dissipative matter
onto a Poincar\'e sphere by defining a Poincar\'e vector
$\overrightarrow{n}(t) = \tr(\rho(t)\overrightarrow{\sigma})$, where the three components are given by
\begin{eqnarray}
u(t)=\rho_{12}(t) + \rho_{21}(t),
\end{eqnarray}
which is a real part of the overlaps between the neutrinos $\nu_e$
and $\nu_\mu$, measuring a reflection between the two neutrinos,
\begin{eqnarray}
v(t)=i(\rho_{12}(t) - \rho_{21}(t)),
\end{eqnarray}
which is an imaginary part of the overlaps, parameterizing an
absorption between the two neutrinos, and
\begin{eqnarray}
w(t)=\rho_{11}(t) -
\rho_{22}(t),
\end{eqnarray}
which describes the transition $\nu_e\rightarrow \nu_\mu$ in process of the
neutrino propagation.

The dynamics of neutrino oscillations described by the
master equation is qualitatively converted into the Poincar\'e
picture in terms of matrix form, i.e.,
\begin{eqnarray}
\frac{d}{dt}\left(\begin{array}{c} u(t)
\\v(t)\\w(t)\end{array}\right) = \left(\begin{array}{ccc}
-2\Gamma_{23} & -2{\cal B}_- & 2{\cal D}_+\\2{\cal B}_+ & -2\Gamma_{13} & -2{\cal C}_-
\\-2{\cal D}_- & 2{\cal C}_+ & -2\Gamma_{12}\end{array}\right)\left(\begin{array}{c} u(t)\\v(t)\\w(t)\end{array}\right),
\end{eqnarray}
where $\Gamma_{ij} = c_{ii} + c_{jj}$, ${\cal
B}_+=\frac{V_0}{2}\cos2\theta -\frac{\Delta
m^2_{21}}{4E}+c_{12},{\cal B}_-=\frac{V_0}{2}\cos2\theta
-\frac{\Delta m^2_{21}}{4E}-c_{21},{\cal
C}_+=\frac{V_0}{2}\sin2\theta\cos\phi+c_{23},{\cal
C}_-=\frac{V_0}{2}\sin2\theta\cos\phi-c_{32}, {\cal
D}_+=\frac{V_0}{2}\sin2\theta\sin\phi+c_{31}$, and ${\cal
D}_-=\frac{V_0}{2}\sin2\theta\sin\phi-c_{13}$. Eq. (22) is called as
a Poincar\'e equation of neutrino oscillations. In order to get its
analytic solution \cite{29}, we firstly diagonalize the $3\times 3$ matrix in
Eq. (22). The three diagonal elements are given by
\begin{eqnarray}
\lambda_0 &=& - \frac{4}{3}\Gamma -
\frac{2^{1/3}}{3}\frac{b}{(a+\sqrt{4b^3+a^2})^{1/3}} +
\frac{1}{3}\left(\frac{a+\sqrt{4b^3+a^2}}{2}\right)^{1/3},
\end{eqnarray}
and
\begin{eqnarray}
\lambda_\pm &=& - \frac{4}{3}\Gamma + \frac{1\pm
i\sqrt{3}}{3}\frac{b}{[4(a+\sqrt{4b^3+a^2})]^{1/3}} -
\frac{1\mp
i\sqrt{3}}{6}\left(\frac{a+\sqrt{4b^3+a^2}}{2}\right)^{1/3},
\end{eqnarray}
with $a = 8(27{\cal B}_+{\cal C}_+{\cal D}_+ + 9 {\cal C}_+{\cal
C}_-(\Gamma_{12}+\Gamma_{13}-2\Gamma_{23}) + (9{\cal D}_+{\cal
D}_--(\Gamma_{12}+\Gamma_{13}-2\Gamma_{23})(2\Gamma_{12}-\Gamma_{13}-\Gamma_{23}))(\Gamma_{12}-2\Gamma_{13}+\Gamma_{23})
+ 9{\cal B}_-(-3{\cal C}_-{\cal D}_- + {\cal
B}_+(-2\Gamma_{12}+\Gamma_{13}+\Gamma_{23})))$ and $b = - 4\Gamma^2
+ 12({\cal B}_+{\cal B}_- + {\cal C}_+{\cal C}_- + {\cal D}_+{\cal
D}_- + \Gamma_{13}\Gamma_{23} +
\Gamma_{12}(\Gamma_{13}+\Gamma_{23}))$.

In terms of the diagonal elements, i.e., $\lambda_0$ and
$\lambda_\pm$,  the solution of Poincar\'e equation (22) can be written
as
\begin{eqnarray}
\left(\begin{array}{c} u(t)\\v(t)\\w(t)\end{array}\right)
=\sum_{i=0,\pm}d_ie^{\lambda_it}\left(\begin{array}{c} 4{\cal
C}_+{\cal C}_- + \Lambda_i\Xi_i\\4{\cal C}_-{\cal D}_ - + 2{\cal
B}_+\Lambda_i\\4{\cal B}_+{\cal C}_ + - 2{\cal
D}_-\Xi_i\end{array}\right),
\end{eqnarray}
where $\Lambda_i = \lambda_i + 2\Gamma_{12}$ and $\Xi_i=\lambda_i +
2\Gamma_{13}$. The time-independent constants
$d_i(\lambda_0,\lambda_+,\lambda_-)$ are determined by the initial
conditions, i.e., $u(0)=\sin(2\theta)\cos\phi$,
$v(0)=\sin(2\theta)\sin\phi$ and $w(0)=\cos(2\theta)$ from the
initial density $\rho_{e}(0)$. We find that
$d_0(\lambda_0,\lambda_+,\lambda_-)=(4u(0){\cal B}_+^2{\cal C}_+ +
v(0){\cal D}_-\Xi_2\Xi_3 + {\cal B}_+(4u(0){\cal
D}_-(\Gamma_{12}-\Gamma_{13}) + w(0)\Lambda_2\Lambda_3 -2v(0){\cal
C}_+(\Lambda_2 + \Xi_3)) + 2{\cal C}_-(4u(0){\cal D}_-^2 -
2w(0){\cal B}_+{\cal C}_+ + {\cal D}_-(-2v(0){\cal C}_+ +
w(0)((\Lambda_2 + \Xi_3))))/(4({\cal C}_-{\cal D}_-^2 + {\cal
B}_+({\cal B}_+{\cal C}_+ + {\cal
D}_-(\Gamma_{12}-\Gamma_{13})))(\lambda_0-\lambda_+)(\lambda_0-\lambda_-))$,
$d_+(\lambda_0,\lambda_+,\lambda_-)=d_0(\lambda_+,\lambda_-,\lambda_0)$
and
$d_-(\lambda_0,\lambda_+,\lambda_-)=d_0(\lambda_-,\lambda_0,\lambda_+)$. The complex numbers $\lambda_\pm$ are very helpful to investigate the neutrino propagation as shown in Ref.\cite{2}.

The radius of Poincar\'e sphere is defined by
\begin{eqnarray}
r^2(t)=\overrightarrow{n}(t)\cdot\overrightarrow{n}(t)=u^2(t)+v^2(t)+w^2(t),
\end{eqnarray}
which is a geometric physical quantity with the effects of reflection form $u(t) = \rho_{12}(t) + \rho_{21}(t)$, absorption from $v(t) = i(\rho_{12}(t) - \rho_{21}(t))$ and transition $w(t) = \rho_{11}(t) -
\rho_{22}(t)$ between the neutrinos $\nu_e$ and $\nu_\mu$ in process of the neutrino mixing and oscillations. Such a radius carries with almost messages of the neutrino mixing and oscillations and therefore provides an important clue to investigate the CPT violating effect in the neutrino mixing and oscillations. Two azimuthal angles can be defined as
\begin{eqnarray}
\alpha(t)=\cos^{-1}\frac{w(t)}{r(t)}, \beta(t)=\tan^{-1}\frac{v(t)}{u(t)}.
\end{eqnarray}

In the Poincar\'e sphere representation, thus, the Poincar\'e vector is parameterized as
\begin{eqnarray}
\overrightarrow{n}(t) = (r(t)\sin\alpha(t)\cos\beta(t), r(t)\sin\alpha(t)\sin\beta(t), r(t)\cos\alpha(t)),
\end{eqnarray}
and density matrix $\rho(t)$ can be expressed by
\begin{eqnarray}
\rho(t) = \frac{1}{2}(1 + \overrightarrow{n}(t)\cdot\overrightarrow{\sigma}),
\end{eqnarray}
with two eigenstates,

\begin{eqnarray}
\mid\nu_e(t)\rangle=\left(\begin{array}{c}
\cos\frac{\alpha(t)}{2}\\ e^{i\beta(t)}\sin\frac{\alpha(t)}{2}
\end{array} \right),
\mid\nu_\mu(t)\rangle=\left(\begin{array}{c}
\sin\frac{\alpha(t)}{2}\\ -e^{i\beta(t)}\cos\frac{\alpha(t)}{2}
\end{array} \right),
\end{eqnarray}
and the corresponding eigenvalues $\lambda_{e}(t)=\frac{1}{2}(1+r(t))$ and $\lambda_{\mu}(t)=\frac{1}{2}(1-r(t))$, respectively. It is obvious that the time-dependent state vectors, $\mid\nu_e(t)\rangle$  and $\mid\nu_\mu(t)\rangle$, are evolving states of
$\mid\nu_e(0)\rangle$ and $\mid\nu_\mu(0)\rangle$, respectively. In the structure of Poincar\'e sphere, the two-flavor neutrino state vectors are two orthogonal antipodal points that lie on the azimuthal angles $\alpha(t)$ and
$\beta(t)$ of the Poincar\'e sphere. Thus the evolution of neutrino system is fully mapped onto the Poincar\'e sphere, where the geometric quantities, i.e., $r(t), \alpha(t)$ and $\beta(t)$, represent the motion trajectory of the two-flavor neutrino mixing and oscillations.

\section{Two-flavor neutrino mixing and CPT-violating mechanism}

Since the eigenvectors of hermitian operator construct a complete
Hilbert subspace, the density matrix $\rho(t)$ can be rewritten as
\begin{eqnarray}
\rho(t)&=&\frac{1}{2}(1+r(t))\mid\nu_e(t)\rangle\langle\nu_e(t)\mid
+ \frac{1}{2}(1-r(t))\mid\nu_\mu(t)\rangle\langle\nu_\mu(t)\mid,
\end{eqnarray}
which indicates that two-flavor neutrino mixed states take $(1\pm r(t))/2$ as the classical mixture probabilities in terms of the radius of Poincar\'e sphere.

The transition probability of neutrinos can be obtained in terms of such a Poinca\'e sphere, i.e.,

\begin{eqnarray}
P(t)&=& \tr\rho(t)\rho_e(0)\nonumber\\&&
=\frac{1}{2}(1+r(t))|\langle\nu_e(0)\mid\nu_e(t)\rangle|^2
+ \frac{1}{2}(1-r(t))|\langle\nu_e(0)\mid\nu_\mu(t)\rangle|^2.
\end{eqnarray}

Inserting Eqs. (17) and (30) into Eq. (32), we find
\begin{eqnarray}
P(t) = \frac{1}{2} + \frac{1}{2}r(t)\left(\cos2\theta\cos\alpha(t) + \sin2\theta\sin\alpha(t)\cos(\beta(t)-\phi)\right),
\end{eqnarray}
which constructs a connection between the geometric quantity (i.e., radius of Poincar\'e sphere) and physical observable (i.e., transition probability). Differently from Eq. (14) in the Schr\"odinger picture, the CP-violating phase $\phi$ is exactly emerged in the transition probability (33) and the radius of Poincar\'e sphere (26). Therefore, it is interesting to observe the dynamic behaviors of the radius of Poincar\'e sphere in terms of the time and CP-violating phase $\phi$.

At the initial time $t=0$, $r(t=0)=1$ leads to
$\rho(t=0)=\rho_{e}(0)$, which includes only a singlet flavor neutrino state. In this case,
the neutrino system is in the pure $\nu_e$ state that can be represented by the surface points on the Poincar\'e sphere. For
an involving state at the time $t>0$, $r(t)<1$ indicates that the density matrix
(31) includes two-flavor neutrino states with the classical mixture
probabilities $0 < (1\pm r(t))/2 < 1$ and therefore the neutrino
system is in a mixed state with both the two-flavor neutrinos $\nu_e$ and $\nu_\mu$. It is obvious
that such two-flavor neutrino mixed states are corresponding to the interior points of
Poincar\'e sphere. When $r(t)=0$, the two-neutrino system is in a maximally mixed state, where the neutrino $\nu_e$ has the same probability $(1\pm r(t))/2=1/2$ as the neutrino $\nu_\mu$.

\begin{figure}
\begin{center}
\epsfig{figure=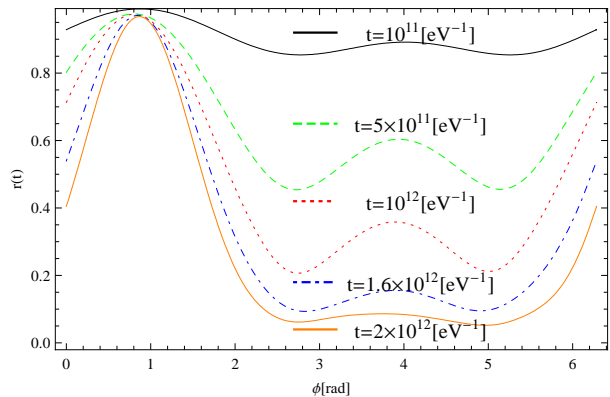,width=8 cm}
\end{center}
\caption{Radius of Poincar\'e sphere as a function of $\phi$ at
different evolving times with the parameters E= 10 MeV, $\Delta
m^2_{21}=8.0\times 10^{-5} eV^2, c_{11}=0.095 V_0, c_{22}=c_{33}=0.15
V_0$, $\theta=0.188\pi$ and $c_{ij}=(c_{ii}c_{jj})^{1/2}$, where
$V_0=\Delta m^2_{21}/2E$, i.e., the matter potential is equal to the neutrino oscillating frequency in vacuum.}
\end{figure}

\begin{figure}
\begin{center}
\epsfig{figure=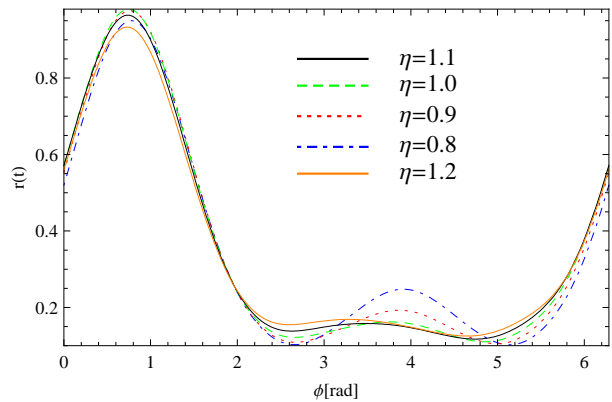,width=8 cm}
\end{center}
\caption{ Radius of Poincar\'e sphere as a function of $\phi$ for
different matter potentials describing by the proportional rate constant $\eta=V_0/(\Delta
m^2_{21}/2E)$ between the matter potential $V_0$ and neutrino oscillating frequency $\omega=\Delta
m^2_{21}/2E$ in vacuum at an evolving time $t=1.9\times 10^{12}
eV^{-1}$ with the parameters E= 10 MeV, $\Delta m^2_{21}=8.0\times
10^{-5} eV^2, \theta=0.188\pi$, $c_{11}=c_{22}=c_{33}=0.1 V_0$ and
$c_{ij}=(c_{ii}c_{jj})^{1/2}$.}
\end{figure}

The radius $r(t)$  illustrates the dynamic characteristics in processing of neutrino mixing and oscillations and defines neutrino mixing degree of freedom in the neutrino oscillations.
It is interesting to show the radius $r(t)$ as a function of time and matter potential in order to know the intrinsically inner evolutions of neutrino propagation. The
Poincar\'e radius as a function of CP-violating angle $\phi$ is
shown in Fig. 1 for $V_0 = \Delta
m^2_{21}/2E$, i.e., the matter potential $V_0$ is equal to the neutrino oscillating frequency $\omega=m^2_{21}/2E$ in vacuum, at different times and in Fig. 2 for different
proportional rate  $\eta = V_0/\omega$ between the matter potential and neutrino oscillating frequency in vacuum for a given time $t=1.9\times
10^{12} eV^{-1}$. Interestingly, we see that beside decreasing
with increasing time because of interaction with the dissipative reservoir, $r(t)$ oscillates in terms of different
amplitudes with an increasing $\phi$, where a big wave peak is emerged in the region of $\phi\in [0,\pi/2]$. In the region of $\phi\in [\pi/2, 2\pi]$, a similar phenomenon of spontaneous symmetry breaking is occurred in the two-flavour neutrino mixing and oscillations, where the "Mexican hat" peak is emerged in the radius of Paincar\'e sphere for the different evolving time (See Fig. 1) and the different matter potential (See Fig. 2).

Generally, spontaneous symmetry breaking is identified with the existence of a non-symmetric lowest energy configuration or state, which accompanies with the dramatic loss of symmetry. The spontaneous symmetry breaking of gauge symmetries is an important component in understanding the origin of particle masses in the standard model of particle physics and the superconductivity of metals. In dynamical gauge symmetry breaking that is a special form of spontaneous symmetry breaking, especially, the bound states of the system itself provide the unstable fields that render the phase transition. Therefore, most phases of matter can be understood through the lens of spontaneous symmetry breaking.

When $t \le 10^{-11} eV^{-1}$ and $t \ge 2\time 10^{-12} eV^{-1}$ as well as $\eta = V_0/(\Delta
m^2_{21}/2E)>1$, the smaller oscillation almost vanishes in the
region $\pi/2\le\phi\le\pi$ (See Figs. 1 and 2). The physical reason may be caused by the Majorana-Dirac confusion theorem \cite{3} because the neutrino masses are not important factor for the neutrino oscillations in these cases.
Such an oscillating behavior of Poincar\'e radius demonstrates that the CP violation enhances the neutrino oscillations just like the matter potential. Therefore, the Poincar\'e radius $r(t)$ of neutrino oscillations may reveal a dynamic mechanism of CPT violation.

\section{Geometric phase}

In situation of pure states, cyclic quantum projections lead to geometric phase which can be understood in terms of geodesic arc on the surface of Poincar\'e sphere \cite{30}. It is known that the pure and mixed states can be represented in the Poincar\'e sphere by a unified way, where the pure states correspond to the surface points on the Poincar\'e sphere and the mixed states are the inside points in the Poincar\'e sphere. Therefore, we try to extend the geometric phase to the mixed state in terms of the Poincar\'e sphere structure.

In order to obtain the geometric phase of mixed state, we firstly replace the state vector \cite{31} by the density matrix and then subdivide the smooth
curve ${\cal C} = \{\rho(t)\}$ into N parts at the points of subdivision $t_0=0, t_1, \cdot\cdot\cdot, t_N = t$, where each trajectory is represented by a discrete
sequence of associated states $\{\rho_0, \rho_1,
\cdot\cdot\cdot, \rho_N,
\rho_{N+1}=\rho_0\}$.

Now let us get the Pancharatnam phase in terms of the density matrix, which is defined as

\begin{eqnarray}
\gamma_{P} &=&
-\arg\tr\lim_{N\rightarrow\infty}\rho_0(t_0)\rho_1(t_1)\rho_2(t_2)\cdot\cdot\cdot\rho_{N-1}(t_{N-1})\rho_N(t_N).
\end{eqnarray}

Under the one order approximation,

\begin{eqnarray}
\langle\sqrt{\lambda_{k_i}}\nu_{k_i}(t_i)
\mid\sqrt{\lambda_{k_{i+1}}(t_{i+1})}\nu_{k_{i+1}}(t_{i+1})\rangle &\approx & \langle\sqrt{\lambda_{k_i}}\nu_{k_i}(t_i)
\mid\sqrt{\lambda_{k_{i+1}}(t_i)}\nu_{k_{i+1}}(t_i)\rangle\nonumber\\&& + \langle\sqrt{\lambda_{k_i}(t_i)}\nu_{k_i}(t_i)
\mid\frac{d}{dt_i}\mid\sqrt{\lambda_{k_{i+1}}(t_i)}\nu_{k_{i+1}}(t_i)\rangle \Delta t_i,
\end{eqnarray}
where $\Delta t_i = t_{i+1} - t_i$, we have

\begin{eqnarray}
\rho_i(t_i)\rho_{i+1}(t_{i+1}) \approx && \sum_{k_i,k_{i+1}}\mid\sqrt{\lambda_{k_i}(t_i)}\nu_{k_i}(t_i)\rangle\langle\sqrt{\lambda_{k_{i+1}}(t_{i+1})}\nu_{k_{i+1}}(t_{i+1})\mid\nonumber\\&&
\times\left(\langle\sqrt{\lambda_{k_i}(t_i)}\nu_{k_i}(t_i)
\mid\sqrt{\lambda_{k_{i+1}}(t_i)}\nu_{k_{i+1}}(t_i)\rangle + \langle\sqrt{\lambda_{k_i}}\nu_{k_i}(t_i)
\mid\frac{d}{dt_i}\mid\sqrt{\lambda_{k_{i+1}}(t_i)}\nu_{k_{i+1}}(t_i)\rangle \Delta t_i\right).
\end{eqnarray}

Inserting Eq. (36) into Eq. (34), the Pancharatnam phase can be expressed as

\begin{eqnarray}
\gamma_{P} &=&
-\arg\tr\lim_{N\rightarrow\infty}\sum_{k_0,k_1,\cdot\cdot\cdot, k_N}\mid\sqrt{\lambda_{k_0}}\nu_{k_0}(t_0)\rangle\langle\sqrt{\lambda_{k_N}}\nu_{k_N}(t_N)\mid\nonumber\\&&
\times\prod_{i=0}^{N}\left(\langle\sqrt{\lambda_{k_i}(t_i)}\nu_{k_i}(t_i)
\mid\sqrt{\lambda_{k_{i+1}}(t_i)}\nu_{k_{i+1}}(t_i)\rangle + \langle\sqrt{\lambda_{k_i}(t_i)}\nu_{k_i}(t_i)
\mid\frac{d}{dt_i}\mid\sqrt{\lambda_{k_{i+1}}(t_i)}\nu_{k_{i+1}}(t_i)\rangle \Delta t_i\right).
\end{eqnarray}

In terms of the adiabatic approximation, we drop off the nondiagonal terms. We find

\begin{eqnarray}
\gamma_{P} &\approx &
-\arg\lim_{N\rightarrow\infty}\langle\sqrt{\lambda_e(t_N)}\nu_e(t_N)\mid\sqrt{\lambda_e(t_0)}\nu_e(t_0)\rangle
\prod_{i=0}^{N}\left(\lambda_e(t_i) + \langle\sqrt{\lambda_e(t_i)}\nu_e(t_i)
\mid\frac{d}{dt_i}\mid\sqrt{\lambda_e(t_i)}\nu_{e}(t_i)\rangle \Delta t_i\right)\nonumber\\&&
-\arg\lim_{N\rightarrow\infty}\langle\sqrt{\lambda_\mu(t_N)}\nu_e(t_N)\mid\sqrt{\lambda_\mu(t_0)}\nu_e(t_0)\rangle
\prod_{i=0}^{N}\left(\lambda_\mu(t_i) + \langle\sqrt{\lambda_\mu(t_i)}\nu_\mu(t_i)
\mid\frac{d}{dt_i}\mid\sqrt{\lambda_\mu(t_i)}\nu_{\mu}(t_i)\rangle \Delta t_i\right)\nonumber\\&&
\approx
-\arg\lim_{N\rightarrow\infty}\langle\sqrt{\lambda_e(t_N)}\nu_e(t_N)\mid\sqrt{\lambda_e(t_0)}\nu_e(t_0)\rangle
\exp\sum_{i=0}^{N}\left(- \lambda_\mu(t_i) + \langle\sqrt{\lambda_e(t_i)}\nu_e(t_i)
\mid\frac{d}{dt_i}\mid\sqrt{\lambda_e(t_i)}\nu_{e}(t_i)\rangle \Delta t_i\right)\nonumber\\&&
-\arg\lim_{N\rightarrow\infty}\langle\sqrt{\lambda_\mu(t_N)}\nu_e(t_N)\mid\sqrt{\lambda_\mu(t_0)}\nu_e(t_0)\rangle
\exp\sum_{i=0}^{N}\left(- \lambda_e(t_i) + \langle\sqrt{\lambda_\mu(t_i)}\nu_\mu(t_i)
\mid\frac{d}{dt_i}\mid\sqrt{\lambda_\mu(t_i)}\nu_{\mu}(t_i)\rangle \Delta t_i\right)\nonumber\\&&
=\arg\sum_{i=e,\mu}\left(\sqrt{\lambda_i(0)\lambda_i(t)}\langle\nu_i(0)\mid\nu_i(t)\rangle\right)
-\Im\sum_{i=e,\mu}\int^{t}_{0}\lambda_i(t)\langle\nu_i(t)\mid
\frac{d}{dt}\mid\nu_i(t)\rangle dt,
\end{eqnarray}
where is invariant under the $U(1)$ gauge transformation,
\begin{eqnarray}
\mid\nu_i(t)\rangle\rightarrow \mid\nu'_i(t)\rangle = e^{i\varphi(t)}\mid\nu_i(t)\rangle, (i=e, \mu),
\end{eqnarray}
with the arbitrary phase factors $\varphi(t)$. For the two-level system, the Pacar\'e sphere, together with an overall $U(1)$ phase, provides a complete SU(2) description \cite{{32},{33}}.

In comparison with the density operator (31) with a $U(1)\times U(1)$ invariant, we see that Eq. (38) includes the contributions of two states and their distribution probabilities in the density matrix but keeps only the U(1) invariant similar to the gauge fixing \cite{{34},{35}}, i.e., $\varphi_e(t)=\varphi_\mu(t)=\varphi(t)$. The geometric phase under the gauge fixing \cite{36} was verified for an incoherent average of pure state interference in terms of the nuclear magnetic resonance technique \cite{37}.

In Eq. (38), the first term on the right hand provides a message of phase differences between the initial and final eigenstates of density matrix and therefore is called as a total phase. At the initial time $r(t=0)=1$, the two classical mixed probabilities $\lambda_{\mu}(0)=(1-r(0))/2=0$ and $\lambda_{e}(0)=(1+r(0))/2=1\ne 0$. Thus
the evolving state of flavor neutrino $\nu_\mu$ does not contribute
to such a total phase. The total phase can be expressed in terms of the
Poincar\'e parameters, i.e.,
\begin{widetext}
\begin{eqnarray}
\gamma_{t}=\tan^{-1}\frac{\sin(\beta(t)-\beta(0))
\sin{\frac{\alpha(0)}{2}\sin{\frac{\alpha(t)}{2}}}}
{\cos{\frac{\alpha(0)}{2}}\cos{\frac{\alpha(t)}{2}}+\cos(\beta(t)-\beta(0))
\sin{\frac{\alpha(0)}{2}\sin{\frac{\alpha(t)}{2}}}}
.
\end{eqnarray}
\end{widetext}
The second term depends on the dynamic evolution of eigenstates of density matrix and therefore is called as dynamic phase and can be separated into two parts. The dynamic phase $\gamma_{d1}$ from the neutrino $\nu_e$ is given by
\begin{eqnarray}
\gamma_{d1}=
-\frac{1}{2}\int_{0}^{t}(1+r(t))\sin^{2}{\frac{\alpha(t)}{2}}d\beta(t),
\end{eqnarray}
and $\gamma_{d2}$ from the neutrino $\nu_\mu$ is
\begin{eqnarray}
\gamma_{d2}=-\frac{1}{2}\int_{0}^{t}(1-r(t))\cos^{2}{\frac{\alpha(t)}{2}}d\beta(t).
\end{eqnarray}

It is known that the dynamic phase of pure state can be extracted quite easily in Hamiltonian formulation. Generally, a mixed state is expressed in many different expansions as a classical probabilistic mixture of pure states. Therefore, it is necessary to consider the classical probability distributions (i.e., the eigenvalues of density matrix) in case of mixed state. Under the transformation,
\begin{eqnarray}
\mid \widetilde{\nu}_i(t)\rangle = \mid \sqrt{\lambda_i(t)}\nu_i(t)\rangle\exp\left(-\int^{t}_{0}\lambda_i(t)\langle\nu_i(t)\mid
\frac{d}{dt}\mid\nu_i(t)\rangle dt\right), (i=e,\mu),
\end{eqnarray}
the Pancharatnam phase (38) can be reexpressed as
\begin{eqnarray}
\gamma_{P} = \arg\sum_{i=e,\mu}\langle\widetilde{\nu}_i(0)\mid \widetilde{\nu}_i(t)\rangle,
\end{eqnarray}
which extracts the dynamic phases of mixed state.

\section{Discussions}

The Pancharatnam phases as a function of CP-violating angle $\phi$ are shown in Figs. 3-5 at different evolving
times under the fluctuational fields. We find that the
Pancharatnam phases are an oscillating function of
$\phi$ and related to the evolving time. The different
curve shapes of Pancharatnam phases show that the time reversal
is not invariant, in turn, the CPT symmetry is violated. The
results provide a useful tool to measure the neutrino property in
the experiments in terms of the time-dependent geometric phase.

\begin{figure}
\begin{center}
\epsfig{figure=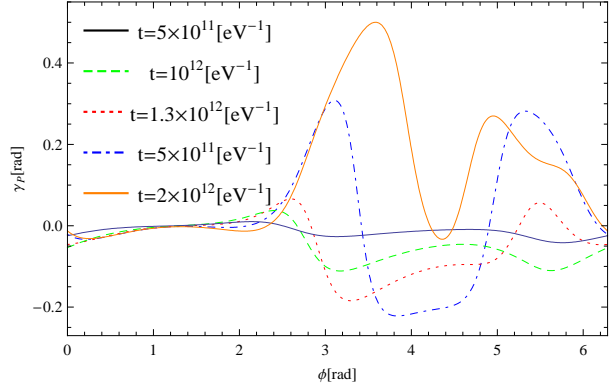,width=8 cm}
\end{center}
\caption{ Pancharatnam phase as a function of $\phi$ at different
evolving times with the parameters E= 10 MeV, $\Delta
m^2_{21}=8.0\times 10^{-5} eV^2, \theta=0.188\pi$, $c_{11}=0.095 V_0,
c_{22}=c_{33}=0.15 V_0$, $c_{ij}=(c_{ii}c_{jj})^{1/2}$ and
$\eta=1$.}
\end{figure}

\begin{figure}
\begin{center}
\epsfig{figure=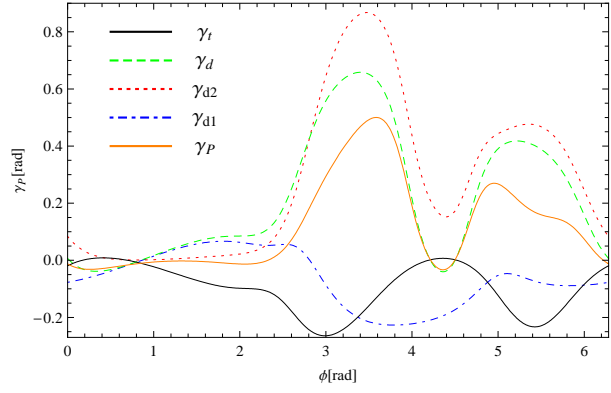,width=8 cm}
\end{center}
\caption{Various different phases in the neutrino mixing and oscillations as a
function of $\phi$ at an evolving time $t=2\times 10^{12} eV^{-1}$
with the parameters E= 10 MeV, $\Delta m^2_{21}=8.0\times 10^{-5} eV^2,
\theta=0.188\pi$, $c_{11}=0.095 V_0, c_{22}=c_{33}=0.15 V_0$,
$c_{ij}=(c_{ii}c_{jj})^{1/2}$ and $\eta=1$, where $\gamma_{d1}$ and $\gamma_{d2}$ are dynamic phases of the neutrinos $\nu_e$ and $\nu_\mu$, respectively. $\gamma_{d}=\gamma_{d1} + \gamma_{d2}$ is a total dynamic phase, $\gamma_{t}$ is a total phase and $\gamma_{P}$ is a Pancharatnam phase.}
\end{figure}

\begin{figure}
\begin{center}
\epsfig{figure=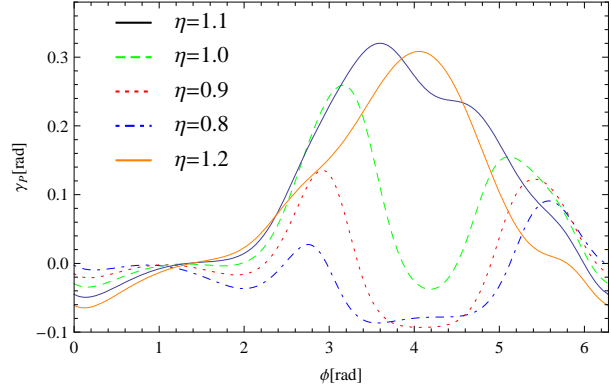,width=8 cm}
\end{center}
\caption{Pancharatnam phase as a function of $\phi$ for different
matter potentials at an evolving time $t=1.9\times 10^{12} eV^{-1}$
with the parameters E= 10 MeV, $\Delta m^2_{21}=8.0\times 10^{-5} eV^2,
\theta=0.188\pi$, $c_{11}=c_{22}=c_{33}=0.1 V_0$ and
$c_{ij}=(c_{ii}c_{jj})^{1/2}$.}
\end{figure}

From Fig. 3, we see that the Pancharatnam phases are small negative
values in the region of $0\le \phi\le 4\pi/5$ and show almost the same
behavior for different evolving times. In the other regions, however,
the oscillations of Pancharatnam phases are different and
dependent on the evolving times. These different behaviors are
resulted from the mixing effects of both the flavor neutrinos $\nu_e$ and
$\nu_\mu$. The contributions of the neutrinos $\nu_e$ and $\nu_\mu$ to the Pancharatnam phases are shown
in Fig. 4. We find that the evolving ways of the dynamic phase
$\gamma_{d1}$ of the neutrino $\nu_e$ are different from
$\gamma_{d2}$ of the neutrino $\nu_\mu$, where $\gamma_{d1}$ is a two-peak structure in the region of $\phi\in [4\pi/5, 2\pi]$ differently from the wave trough of $\gamma_{d2}$ and the iterating results of
both $\gamma_{d1}$ and $\gamma_{d2}$ lead to the total dynamic
phase $\gamma_{d}=\gamma_{d1} + \gamma_{d1}$. Therefore, the mixing
effects of neutrinos are very important to the geometric phase
because the neutrino system is in a superposition with the
$\nu_e$ and $\nu_\mu$. Our results are different from Ref. \cite{20}, where
the kinematic approach was used and the mixture probabilities were taken
as the constants $\sqrt{\lambda_k(t=0)\lambda_k(t=T)}$ (T is a
quasicyclity, $k=e, \mu$). We know that $\lambda_{\mu}(t=0)= (1-
r(t=0))/2=0$ and therefore the contribution of the neutrino $\nu_\mu$ to
the geometric phase vanishes in all evolving processes. Thus the kinematic
approach to geometric phase does not include the mixing effect.

The Pancharatnam phases of neutrino oscillations depend also on the
matter potential as shown in Fig. 5. When the matter potential $V_0$
is larger than the neutrino oscillating frequency $\omega = \Delta
m^2_{21}/(2E)$ in vacuum, i.e., $\eta >1$, the two peaks structures ($\eta <1$)
become single peak.

\section{Detections}

It is well-known that the conventional way to address the Majorana issue given the small mass scale of the neutrino in this field is neutrinoless double beta decay (NDBD). Because of the extremely low efficiency of direct neutrino measurements in terms of the physical observable (i.e., transition probability of neutrino), unfortunately, it is necessary to accomplish  exquisitely complex state-of-the-art rare event infrastructure.

Another observable is geometric phase in the neutrino mixing and oscillations. Thus the time-dependent Pancharatnam phase can be helpful in the detection of CPT-violation due to an enhanced
flexibility in the choice of evolutions as shown in Figs. 3-5. This phase factor can be realized by
the neutrino split-beam-interference experiment \cite{10}. Unfortunately, the
spatially beam splitting experiment is almost impossible for a tiny
interaction cross section. When the neutrino propagates
through dissipative matter, on the other hand, the environment is usually unknown. Thus it is difficult to
determine the decay coefficients $c_{ij}$ in the neutrino experiments.

A possible alternative approach is quantum simulation, where the simulation of fundamental particle physics properties using a controllable quantum system is an increasing and interesting topic in both particle physics and condense matter physics \cite{{37},{38},{39},{40},{41}}. Thus we propose to detect the Pancharatnam phase by simulating the
environment of neutrino propagation in terms of a
controllable physical field (e.g., magnetic field), ${\cal M}_0 = \frac{1}{2}V_0\sigma_z$,
with corresponding fluctuation fields ${\cal M}_i = d_i\sigma_i$ (e.g., fluctuational magnetic fields), and a
nuclear-magnetic-resonance (NMR) system with the Hamiltonian \cite{42},
\begin{eqnarray}
{\cal
H}_{NMR} =-\frac{1}{2}\omega\cos2\theta\sigma_z +
\frac{1}{2}\omega\sin2\theta\cos\phi\sigma_x +
\frac{1}{2}\omega\sin2\theta\sin\phi\sigma_y,
\end{eqnarray}
which has the same form as Eq. (1) beside the firm term for the two-flavor neutrino oscillations in medium and
the simulating parameters are taken in terms of the parameters from the neutrino oscillations, i.e.,  $\omega=\Delta
m^2_{21}/2E\in [4.0\times 10^{-12}, 4.0\times 10^{-11}] eV^{-1}, V_0=\eta\Delta
m^2_{21}/2E\in [4\eta\times 10^{-12}, 4\eta\times 10^{-11}] eV^{-1}$ and $d_i=0.1V_0\in [4\eta\times 10^{-13}, 4\eta\times 10^{-12}] eV^{-1}$ for the neutrino square-mass difference $\Delta m^2_{12}=8.0\times 10^{-5} eV^2$ and energy range $E\in [1, 10]$MeV with a controlling and adjusting constant $\eta$ as shown in Figs. 1-5. Thus the Hamiltonian (1) of neutrino oscillations is
converted equivalently into ${\cal H}={\cal M}_0 + {\cal
H}_{NMR}$, where the constant term ${\cal H}_0$ in Eq. (1) is dropped off because it
only contributes an overall phase factor. Next, using the relation $[\sigma_i\rho, \sigma_j] +
[\sigma_i,\rho \sigma_j]=-[\sigma_i, [\sigma_j, \rho]]$, the
dissipative terms (16) in the master equation (15) can be reexpressed by
\begin{eqnarray}
{\cal L}\rho = -\frac{1}{2}\sum_{i,j=x,y,z}[{\cal M}_i, [{\cal M}_j,
\rho]],
\end{eqnarray}
where the double commutator
results in a time-irreversibility and is helpful to study the
CPT violation. Thus the environment effects can also simulated by the controlable fluctuational fields. Comparing Eq. (46) with Eq. (16), we take $c_{ij}=d_id_j$ as the decay coefficients.

In terms of Eqs. (45)-(46), we construct a connection between the neutrino oscillation and NMR experiment. It was demonstrated in
the NMR experiment that an open system can be simulated by varying
the choice of mapping between the simulated system and the simulator
\cite{37}.  An important advantage of quantum simulations
\cite{{38},{39},{40},{41}} is that the values of the neutrino mass square difference, mixing angle of vacuum, propagating energy,
CP-violating angle and dissipative coefficients
 can be measured just by controlling the fluctuational field and NMR parameters. On the other hand, it becomes possible and easy to detect the Pancharatnam phase of neutrino
through dissipative matter in terms of the controllable quantum system, i.e., NMR system.

\section{Conclusions}

In summary, the neutrino mixing and oscillations are modeled on the basis of a
two-dimensional Hilbert space with an intrinsic CP-violating phase under the dissipative matter. The geometric representation is given by mapping the dynamic
master equation onto the Poincar\'e sphere structure. We show
that the mixture of two-flavor neutrino system is
in terms of the Poincar\'e radius depending on the intrinsic CP-violating phase. We find that, especially, the phenomena of both similar spontaneous symmetry breaking mechanism and Majorana-Dirac neutrino confusion are emerged in such a time-dependent Poincar\'e radius from the neutrino mixing and oscillations. Moreover, we find that  the CPT-violating effect is enhanced in the time-dependent
geometric phase under the neutrino propagating process. The results provide a new way to test the Dirac
vs Majorana neutrinos and Majorana-Dirac confusion theorem as well as observe the CPT-violation. At last, we propose to detect such a geometric phase in terms of the quantum simulation by using a controllable physical field and NMR system.\\

\noindent\textbf{Acknowledgments}\\

This work is supported by
the Natural Science Foundation of China under Grant No. 11565015 and
No.11365012, the
Natural Science Foundation of Jiangxi Province, China under Grant
No. 20132BAB202008, the Foundation of Science and Technology of
Education Office of Jiangxi Province under No. GJJ13235.

Hui Pan thanks the supports of the Science and Technology Development Fund from Macao
SAR (FDCT-068/2014/A2 and FDCT-132/2014/A3), and Multi-Year Research Grants
(MYRG2014-00159-FST and MYRG2015-0015-FST) and Start-up Research Grant (SRG-
2013-00033-FST) from Research and Development Office at University of Macau.

\end{document}